\documentclass[preprints,article,accept,moreauthors,pdftex]{mdpi}

\firstpage{1} 
\pubvolume{xx}
\issuenum{1}
\articlenumber{5}
\pubyear{2019}
\copyrightyear{2019}
\history{Received: date; Accepted: date; Published: date}

\Title{A Fast Universal Kinematic Fitting code for low-energy nuclear physics: \texttt{FUNKI\_FIT}}

\Author{Robin Smith $^{1,2,\dagger}$\orcidA{} and Jack Bishop $^{3,\dagger}$\orcidB{}}

\AuthorNames{Robin Smith, Jack Bishop}

\address{%
$^{1}$ \quad Faculty of Science, Technology \& Art, Sheffield Hallam University, Sheffield, S1 1WB, UK; robin.smith@shu.ac.uk\\
$^{2}$ \quad The LNS at Avery Point, University of Connecticut, Groton, CT 06340-6097, USA\\
$^{3}$ \quad Department of Physics \& Astronomy and Cyclotron Institute, Texas A\&M University, College Station, TX 77843, USA; jackbishop@tamu.edu}

\firstnote{These authors contributed equally to this work.}

\abstract{We present an open source kinematic fitting routine designed for low-energy nuclear physics applications. Although kinematic fitting is commonly used in high-energy particle physics, it is rarely used in low-energy nuclear physics, despite its effectiveness. A \texttt{FORTRAN} and \texttt{ROOT C++} version of the \texttt{FUNKI\_FIT} kinematic fitting code have been developed and published open access. The \texttt{FUNKI\_FIT} code is \emph{universal} in the sense that the constraint equations can be easily modified to suit different experimental set-ups and reactions. Two case studies for the use of this code, utilising experimental and Monte-Carlo data, are presented: (1) charged-particle spectroscopy using silicon-strip detectors; (2) charged-particle spectroscopy using active target detectors. The kinematic fitting routine provides an improvement in resolution in both cases, demonstrating, for the first time, the applicability of kinematic fitting across a range of nuclear physics applications. The \texttt{ROOT} macro has been developed in order to easily apply this technique in standard data analysis routines used by the nuclear physics community.
}


\begin{document}


\section{Introduction}
In all nuclear physics experiments, measurements are subject to various uncertainties, both statistical and systematic. A typical experiment involves a beam of accelerated sub-atomic particles impinging on a target. Upon a nuclear reaction taking place, radiation is emitted. Such experiments involve measuring this radiation (charged ions, electrons, neutrons or electromagnetic radiation) and using it to reconstruct the reaction taking place. Charged ions may be detected using silicon-strip detectors or gaseous Time Projection Chambers (TPCs) \cite{Pagano(2016),Gai(2019),TexAT1}. Such devices provide information about the \emph{energies} and \emph{directions} of the particles they detect.\\


Depending on the experiment, various uncertainties limit the ability to reconstruct these reactions. These uncertainties can arise from the beam of particles itself. Modern, radioactive beam experiments can provide intense beams of radioactive isotopes in order to study the behaviour of nuclei far from stability \cite{Catherall(2017),Bark(2018)}. However, the composition of the beam, its energy and spatial location are sometimes poorly constrained. Other experimental uncertainties arise from radiation detection. The detector type and material defines the energy resolution, and the physical size of the detector can limit the spatial resolution (how well the direction of a particle can be determined). Due to these experimental limitations, the momenta of the particles reconstructed in the final state, and hence other information about the nuclear reaction, cannot be known exactly. However, it is possible to significantly improve this resolution of the measured quantities by applying a process called \emph{kinematic fitting}, which is widely used in particle physics \cite{KinFitPP} but has rarely been used in nuclear spectroscopy \cite{KirsebomHoyle,RobinThesis}.\\

The idea of kinematic fitting is to identify and use physical kinematic constraints of the experimental system, and vary the experimental measurements (energies and directions of the particles) such that these constraints are matched \emph{exactly}. In theory, this should correct the experimentally measured quantities, bringing them closer to their true values. Two such constraints could be the total energy and vector momentum of a system. If the beam energy and direction is known before a reaction, along with the reaction $Q$-value, the sum of the momenta of the particles after the reaction is constrained, along with the total energy. Such constraints can be used to correct the experimental data. The more constraints that a system has, the more accurately the measured parameters can be known. During the kinematic fitting process, the amount that each parameter can move from its initial value is modulated by its original measurement uncertainty, obtained empirically. This means that poorly measured parameters can change more easily, and parameters with a small uncertainty do not change much from their initial values.\\

Such kinematic fitting routines are commonplace in high-energy physics. They have been used for several decades and are now routinely built into data analysis programs. The present paper clearly demonstrates, for the first time, the usefulness of kinematic fitting in the field of low-energy nuclear physics, across a range of reaction types and detector systems. The aim of this paper is to introduce a kinematic fitting framework, \texttt{FUNKI\_FIT}, which may be utilised by the nuclear physics community to improve the resolution in a variety of data analyses. We initially provide a brief introduction to the mathematics of kinematic fitting using Lagrange multipliers and a simple example is given to demonstrate the method. The \texttt{FUNKI\_FIT} code is then introduced and applied to two important case studies: charged-particle spectroscopy using silicon strip detectors and active target detectors (Time Projection Chambers).

\vspace{0.5cm}


\section{Materials and Methods}

Mathematically, kinematic fitting is performed using the method of Lagrange multipliers, where the difference between the measured parameter values and their true values are minimised subject to a number of constraint equations, defined by the system being measured. Let $\alpha$ be a column vector representing a set of parameters such as the true energies and angles of the particles emerging from a nuclear reaction, (without any experimental uncertainty). If a single event corresponds to measuring $n$ parameters, then $\alpha$ takes the form

\begin{eqnarray}
\alpha = \begin{pmatrix}\alpha_1\\ \alpha_2\\ . \\ . \\ \alpha_n\end{pmatrix}.
\label{eq:AlphaParameters}
\end{eqnarray}

\noindent In an experiment, the parameters are measured as their initial, unconstrained values, contained in the vector, $\alpha_0$. How well the measured parameters $\alpha_0$, which are subject to an experimental resolution $\sigma$, match the true values $\alpha$ can be quantified by computing the value of $\chi^2$ in matrix form as

\begin{eqnarray}
\chi^2 = (\alpha - \alpha_0)V_{\alpha_0}^{-1}(\alpha - \alpha_0),
\label{eq:ChiSqMatrix}
\end{eqnarray}

\noindent where $V_{\alpha_0}^{-1}$ is the inverse of the covariance matrix. In kinematic fitting, the values of the measured parameters are changed so that equation \ref{eq:ChiSqMatrix} is minimised. This minimisation is subject to a number of physical constraints. As an example, if a constraint of the system is that all of the parameters must sum to a constant value, $S$, then the constraint equation is

\begin{eqnarray}
H = \alpha_1 + \alpha_2 \hspace{0.1cm} ... + \alpha_n - S = 0.
\label{eq:ExampleConstraint}
\end{eqnarray}

\noindent If instead, the parameters are subject to a number of $r$ constraint equations, $H_r$, then we wish to minimise $\chi^2$ subject to the constraint that

\begin{eqnarray}
\mathbf{H}(\alpha) = 0.
\label{eq:ConstraintVector}
\end{eqnarray}

\noindent Here, $\mathbf{H}$ is a row vector, which contains each of the constraint equations $(H_1\hspace{0.2cm} H_2 \hspace{0.2cm} . \hspace{0.2cm} . \hspace{0.2cm} . \hspace{0.2cm} H_r)$. Equation \ref{eq:ConstraintVector} can be expanded around an arbitrary point in the parameter space $\alpha_A$ (commonly chosen as $\alpha_A = \alpha_0$) such that

\begin{eqnarray}
\mathbf{H}(\alpha) = 0 \approx H(\alpha_0) + \frac{\partial H(\alpha_A)}{\partial \alpha}(\alpha - \alpha_0) = \mathbf{d} + \mathbf{D} \delta \alpha ,
\label{eq:ConstraintExpansion}
\end{eqnarray}

\noindent where $\delta \alpha = (\alpha - \alpha_0)$ and

\begin{eqnarray}
\mathbf{d} = \begin{pmatrix}H_1(\alpha_A)\\ H_2(\alpha_A)\\ . \\ . \\ H_r(\alpha_A)\end{pmatrix} \hspace{0.5cm} \mathbf{D} =
  \left( {\begin{array}{cccc}
   \frac{\partial H_1}{\partial \alpha_1} & \frac{\partial H_1}{\partial \alpha_2} & \hdots & \frac{\partial H_1}{\partial \alpha_n} \\ \frac{\partial H_2}{\partial \alpha_1} & \frac{\partial H_2}{\partial \alpha_2} & \hdots & \frac{\partial H_2}{\partial \alpha_n} \\ \vdots & \vdots & \ddots & \vdots \\ \frac{\partial H_r}{\partial \alpha_1} & \frac{\partial H_r}{\partial \alpha_2} & \hdots & \frac{\partial H_r}{\partial \alpha_n}\   \end{array} } \right).
\label{eq:ConstraintExpansionMatrices}
\end{eqnarray}

\noindent In \texttt{FUNKI\_FIT}, the $\mathbf{D}$ matrix is obtained by numerically differentiating each constraint equation in the direction of each parameter $\alpha_i$ at the unconstrained $\alpha_0$ values. From here, $\chi^2$ is minimised subject to equation \ref{eq:ConstraintExpansion} using the method of Lagrange multipliers \cite{AveryKinFit} as

\begin{eqnarray}
\chi^2 = (\alpha - \alpha_0)V_{\alpha_0}^{-1}(\alpha - \alpha_0) + 2\lambda^T(\mathbf{d} + \mathbf{D} \delta \alpha).
\label{eq:ConstraintExpansionMatrices}
\end{eqnarray}

\noindent Here, $\lambda$ is a vector of $r$ unknowns. Minimising $\chi^2$ with respect to $\alpha$ and $\lambda$ leads to a solution by solving the two linear vector equations

\begin{eqnarray}
V_{\alpha_0}^{-1}(\alpha - \alpha_0) + \mathbf{D}^T \lambda &=& 0\\
\mathbf{d} + \mathbf{D} \delta \alpha &=& 0.
\label{eq:KinFitSolutions}
\end{eqnarray}

\noindent The solution can be found in steps by computing the following matrix multiplications:

\begin{eqnarray}
\label{eq:KinFitSolutionsSteps1}
V_D &=& (\mathbf{D}V_{\alpha_0}\mathbf{D}^T)^{-1}\\
\label{eq:KinFitSolutionsSteps2}
\lambda &=& V_D(\mathbf{d} + \mathbf{D}\delta \alpha_0).
\end{eqnarray}

\noindent The new, corrected parameters, $\alpha$ may then be computed as

\begin{eqnarray}
\label{eq:KinFitSolutionsSteps3}
\alpha &=& \alpha_0 - V_{\alpha_0}\mathbf{D}^T \lambda .
\end{eqnarray}

\noindent The uncertainties on the new parameters are obtained by calculating the new covariance matrix as

\begin{eqnarray}
V_{\alpha} = V_{\alpha_0} - V_{\alpha_0} \mathbf{D}^T V_D \mathbf{D} V_{\alpha_0} .
\label{eq:CovarianceMatrix}
\end{eqnarray}

\begin{figure}[H]
\centering
\includegraphics[scale=0.55]{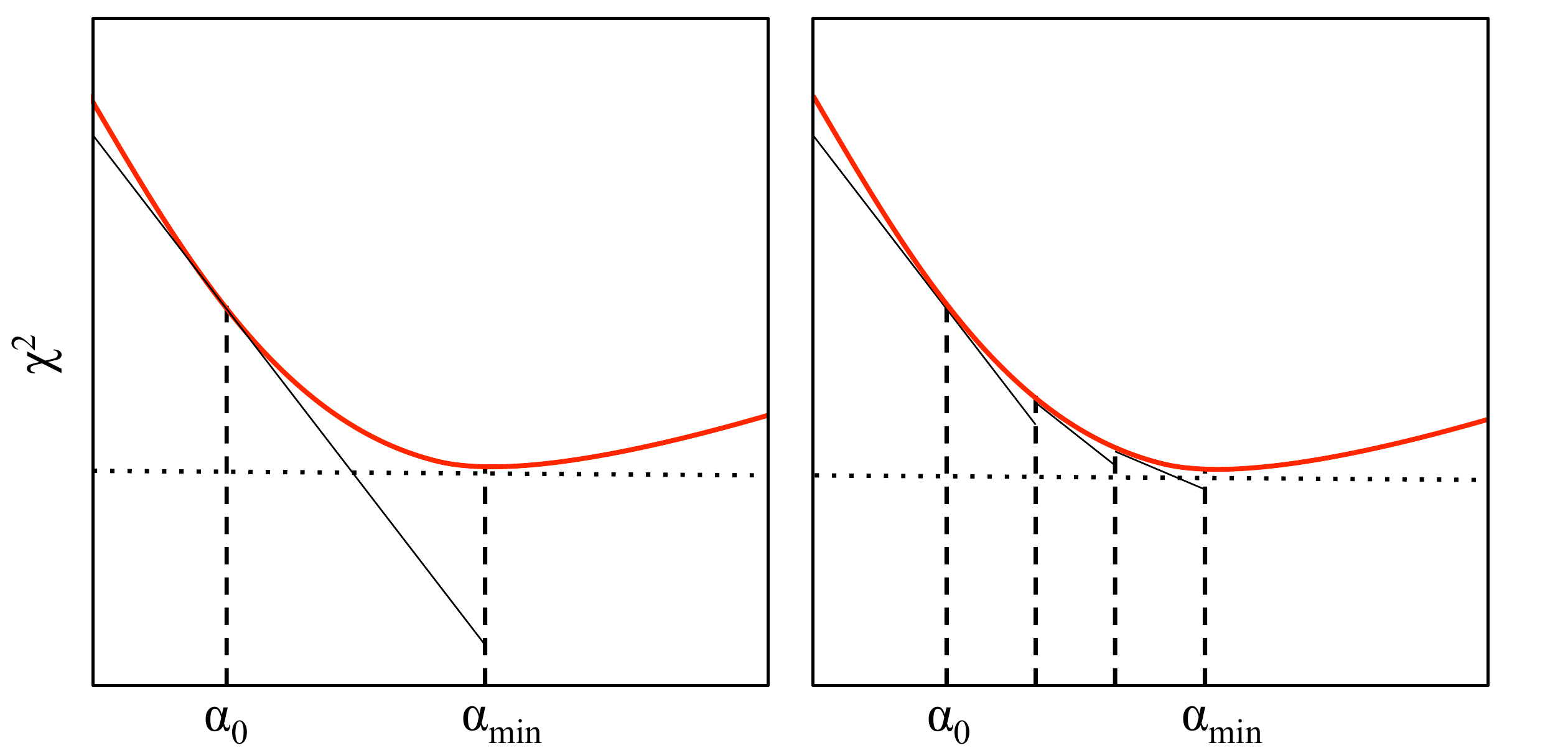}
\caption{Variation of $\chi^2$ due to a hypothetical non-linear constraint equation depending on a parameter, $\alpha$. Left panel: using the prescription of equation \ref{eq:KinFitSolutionsSteps3}, the solution that minimises $\chi^2$ is not found. Right panel: Using the iterative method of equation \ref{eq:Iterations}, the minimum is found.}
\label{fig:convergence}
\end{figure}

\noindent The correction to the parameters, applied by equation \ref{eq:KinFitSolutionsSteps3}, assumes that the constraint equations vary linearly with each of the parameters, $\alpha$. This is because the matrix, $\mathbf{D}$, evaluates the gradient of $\mathbf{H}(\alpha)$ in the direction of each parameter. However, for non-linear constraint equations, as the parameters are moved away from their initial $\alpha_0$ values, the $\mathbf{D}$ matrix will also change. This means that equation \ref{eq:KinFitSolutionsSteps3} will undershoot or overshoot the minimising parameter values. Therefore, in order to find the solution for non-linear constraints, the solution is computed iteratively until the minimum $\chi^2$ is found. Equation \ref{eq:KinFitSolutionsSteps3} is rewritten in the form

\begin{eqnarray}
\alpha &=& \alpha_0 - w V_{\alpha_0}\mathbf{D}^T \lambda ,
\label{eq:Iterations}
\end{eqnarray}

\noindent where $w$ is a weighting factor, which has a value considerably less than 1. A value of 0.01 was chosen in the present work. The kinematic fitting procedure is followed a number of times. Each time, the change in the parameters $\alpha$ is suppressed by $w$. This allows a gradual convergence towards parameter values that minimise $\chi^2$. The procedure is shown pictorially in figure \ref{fig:convergence}.

\begin{figure}[H]
\centering
\includegraphics[scale=0.65]{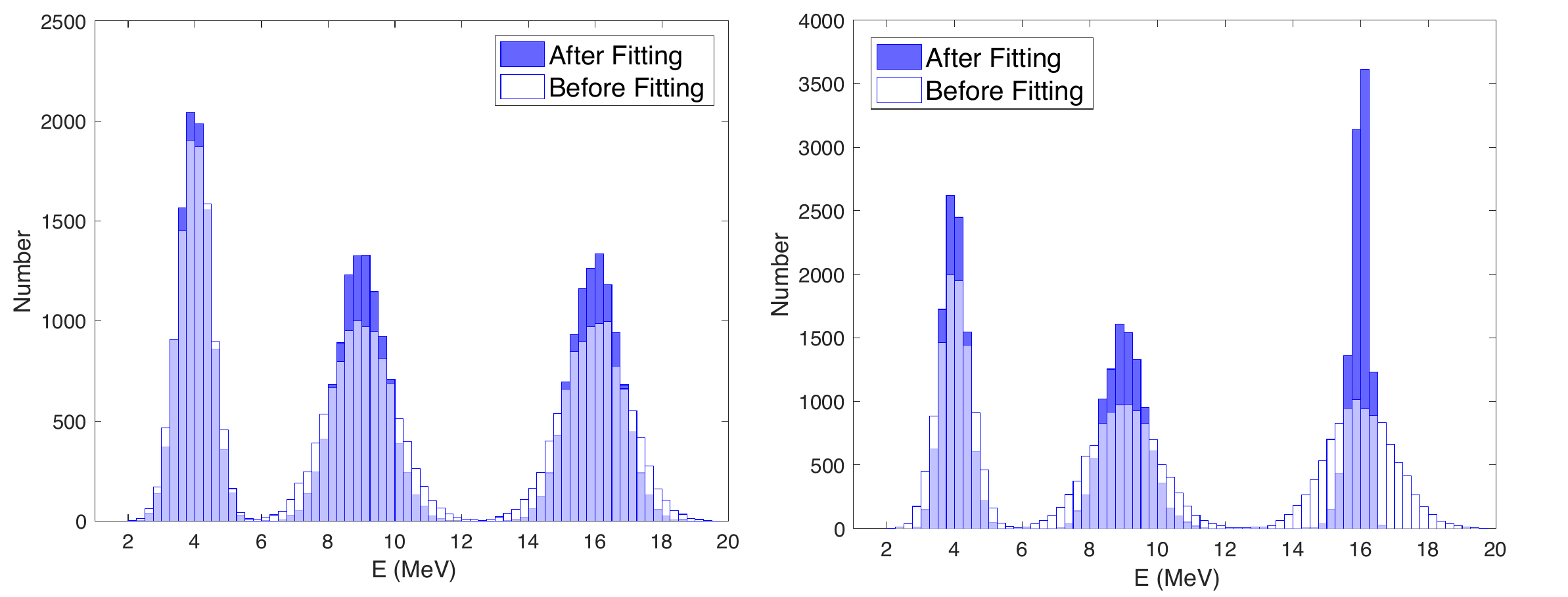}
\caption{The distribution of particle energies before (clear histogram) and after (filled histogram) kinematic fitting. Left panel: One constraint (equation \ref{eq:SimpleConstraints1}). Right Panel: Two constraints (equations \ref{eq:SimpleConstraints1} and \ref{eq:SimpleConstraints2}). Here, $E_3$ is best constrained after fitting since its value has a larger effect on equation \ref{eq:SimpleConstraints2}.}
\label{fig:KinFitEx1}
\end{figure}

To illustrate the power of kinematic fitting, consider the simple example of measuring the energies of three particles ($E_1$,~$E_2$,~$E_3$)~MeV. To simplify the problem, their true values for each measurement are fixed as ($4$,~$9$,~$16$)~MeV. However, the parameters are measured with a resolution ($\sigma_1$,~$\sigma_2$,~$\sigma_3$) = ($0.5$,~$1$,~$1$)~MeV. This simple system has been modelled with a Monte-Carlo simulation. For 10,000 event measurements, the distribution of measured energies are shown in Figure \ref{fig:KinFitEx1} by the clear histogram. It is possible to improve the resolution of these parameters through kinematic fitting with two hypothetical constraint equations. These equations are chosen such that they are satisfied by the true values of the parameters

\begin{eqnarray}
\label{eq:SimpleConstraints1}
E_1 + E_2 + E_3 = E_\textrm{sum} = 29\\
\label{eq:SimpleConstraints2}
E_1^2 + E_2^2 + E_3^2 = E^2 = 353.
\end{eqnarray}

The filled histogram of Figure \ref{fig:KinFitEx1} shows the distribution of measured energies after kinematic fitting with both one and two constraints. The plot is layered such that the distribution after kinematic fitting can be seen through the initial distribution. The uncertainties in the newly-calculated parameters can be easily computed after the fitting process has been performed using equation \ref{eq:CovarianceMatrix}. For the single constraint equation, this gives ($\sigma_1$,~$\sigma_2$,~$\sigma_3$) = ($0.471$,~$0.745$,~$0.745$)~MeV, which agrees with the distributions shown in the left panel of Figure \ref{fig:KinFitEx1}.

The above mathematics for kinematic fitting have been implemented for use in low energy nuclear physics using the Fast Universal Kinematic Fitting (\texttt{FUNKI\_FIT}) code, which has been written for \texttt{FORTRAN} \cite{FUNKIFITfortran} and \texttt{ROOT} C++ \cite{FUNKIFITroot}. In addition to the simple example above, the \texttt{FORTRAN} and C++ codes have been tested on two example data sets, and the results are presented next. The first involves precisely measuring the decay of $^{12}$C into three $\alpha$-particles using a position sensitive silicon detector array. The \texttt{FORTRAN} code has been used for this analysis. Secondly, the $^{10}$C($\alpha$,$\alpha$) resonant elastic scattering reaction was explored using the \texttt{ROOT} C++ code. Detailed Monte-Carlo simulations of the Texas Active Target (TexAT) detector \cite{TexAT1,TexAT2} were used to model this reaction in Thick Target Inverse Kinematics (TTIK). In TTIK, a heavy ion passes through a light gas, allowing the cross section over a range of energies to be measured with a single beam energy.

\vspace{0.5cm}


\section{Results and Discussion}

\subsection{Charged-particle spectroscopy with silicon-strip detectors}

As mentioned in the introduction, it is important to obtain optimum energy and position resolution in charged-particle spectroscopy experiments. This is of utmost importance in experiments where the relative momenta of the detected particles are being examined. This has been seen in several cases relating to borromean nuclei \cite{Marquesdalitzplot,Smith9BeDalitz,SmithHoyle}.\\

\begin{figure}[H]
\centering
\includegraphics[scale=0.6]{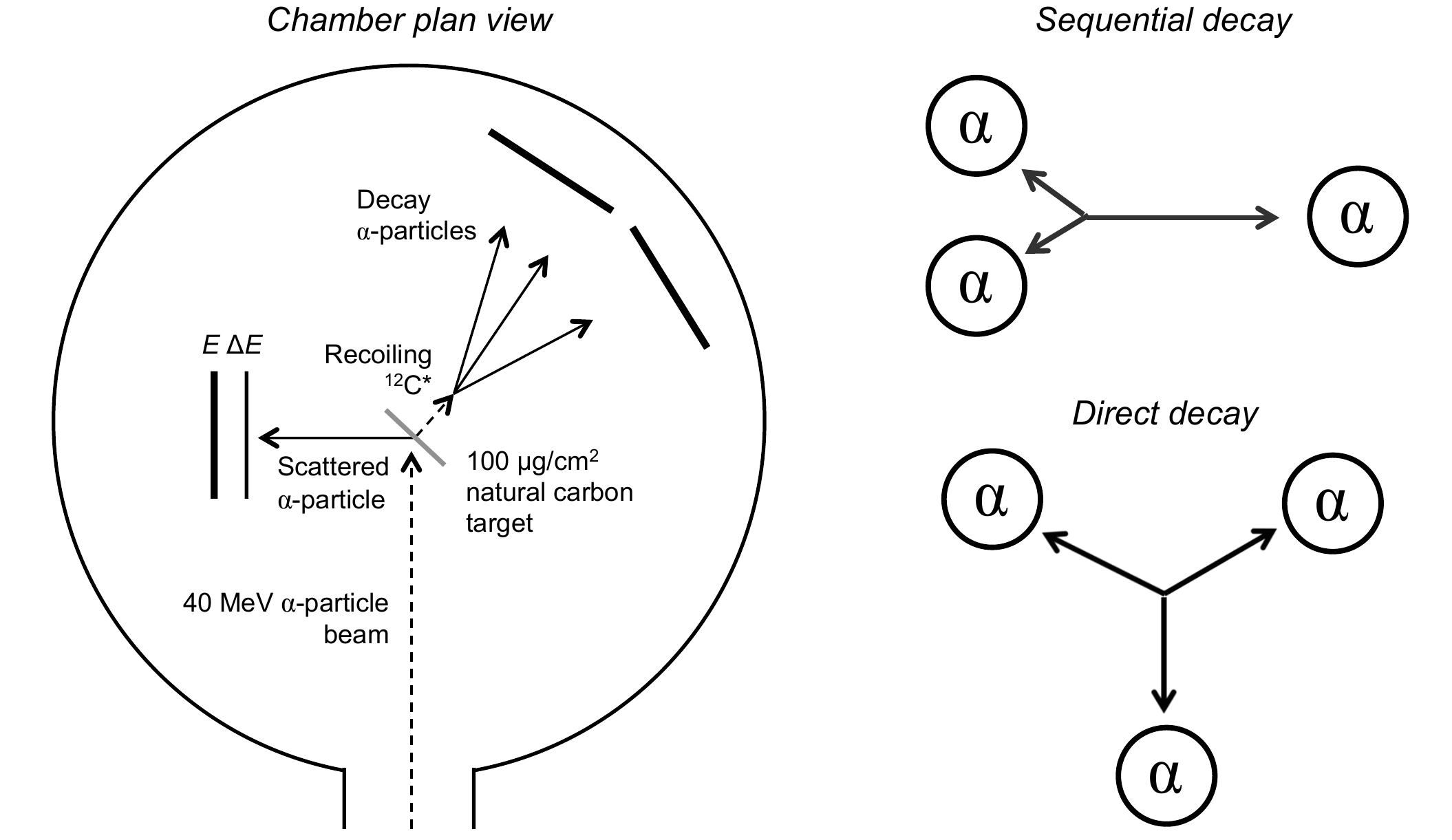}
\caption{Left panel: schematic diagram of the experiment used to measure the 3$\alpha$ break-up of $^{12}$C. Right panel: Pictorial representation of the sequential and direct decays of $^{12}$C.}
\label{fig:ChamberLayout}
\end{figure}

In reference \cite{SmithHoyle}, an upper limit on the branching ratio for direct and sequential decay of the famous Hoyle state in $^{12}$C was measured. This branching ratio is important in astrophysics and for understanding nuclear structure \cite{HoyleReview,AldoHoyle}. The set-up used in the experiment is shown in the left panel of Figure \ref{fig:ChamberLayout}. A beam of 40 MeV $\alpha$-particles, produced by the University of Birmingham MC40 Cyclotron, were directed onto a 100~$\mu$g/cm$^2$ natural carbon target. Inelastic scattering of an $\alpha$-particle from a $^{12}$C nucleus may populate an excited state in carbon. The scattered $\alpha$-particle was detected in a $\Delta E-E$ telescope set-up of two Micron W1 double-sided silicon strip detectors (DSSDs) [Micron Semiconductor Ltd], one 65 $\mu$m-thick and the other 500 $\mu$m-thick, placed at 90$^\circ$ to the beam; this permitted particle identification. The energy of this scattered $\alpha$-particle was used to calculate the excitation energy of the recoiling $^{12}$C, allowing events that populated the Hoyle state to be identified. Then the three other $\alpha$-particles, resulting from the decay of an above-threshold excited state in $^{12}$C, were detected in an array of four 500 $\mu$m-thick Micron W1 DSSDs. It is the relative momenta of these three $\alpha$-particles that allowed the determination of the type of decay.

The right panel of Figure \ref{fig:ChamberLayout} shows a pictorial depiction of the break-up of $^{12}$C into three $\alpha$-particles, in its centre-of-mass frame. In the top \emph{sequential} decay case, the $^{12}$C first emits an $\alpha$-particle with a fixed energy, leaving a recoiling $^8$Be. This $^8$Be is unbound by 92 keV and later decays into two $\alpha$-particles. In the bottom \emph{direct} decay case, where the three $\alpha$-particles are emitted from the nucleus simultaneously, the energy in the decay is shared differently between the three fragments. These two types of decay are disentangled by examining the relative momenta of the three $\alpha$-particles. The measured momenta of the break-up $\alpha$-particles allowed the $^{12}$C recoil momentum to be determined and therefore the energies of the particles in the centre-of-mass of the decaying $^{12}$C could be calculated.\\

In the published work \cite{SmithHoyle}, only events where all three break-up $\alpha$-particles hit the detectors were considered. This is because their momenta were precisely measured and therefore a good resolution was obtained when comparing their relative momenta. However, these type of events make up a small fraction of the total due to incomplete solid angle coverage. A second, much larger, subset of data that were not published consists of those events where one of the break-up $\alpha$-particles misses the detector array entirely. However, since the beam momentum is known, as are the momenta of the three other detected $\alpha$-particles, the vector momentum of the third, undetected $\alpha$-particle may be calculated as\\

\begin{eqnarray}
\vec{P}_{\textrm{$\alpha_4$}} = \vec{P}_{\textrm{beam}} - \sum_{i=1}^{3} \vec{P}_{\alpha_\textrm{i}}.
\label{eq:Undetected}
\end{eqnarray}

However, this does not permit as good resolution as when all four $\alpha$-particles are detected directly. The uncertainties on the energies and polar angles of the three detected $\alpha$-particles are combined in equation \ref{eq:Undetected} meaning that the momentum of the reconstructed $\alpha$-particle is poorly constrained. Therefore, even though higher statistics are possible with this larger data set, the energy resolution is too poor for the data to be useful in analysing the $\alpha$-particle relative momenta.\\

\begin{figure}[H]
\centering
\includegraphics[scale=0.59]{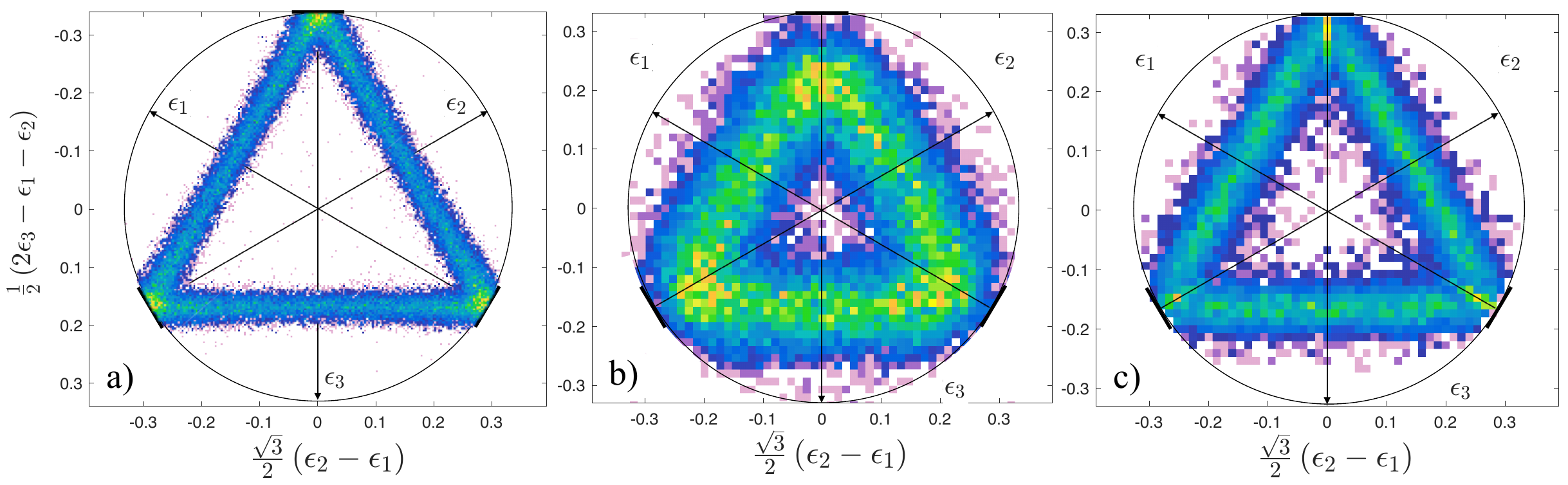}
\caption{Dalitz plots showing the relative fractional energies of $\alpha$-particles emitted from $^{12}$C. Left panel: Complete kinematics data show excellent resolution. Centre panel: The data where one $\alpha$-particle is reconstructed. Right panel: The same data after kinematic fitting was applied.}
\label{fig:ThreeDatitzPlots}
\end{figure}

The relative momenta of the three $\alpha$-particles resulting from the decay of $^{12}$C may be visualised using a Dalitz plot \cite{DalitzPlotOriginal}. Some Dalitz plots from this data set are shown in Figure \ref{fig:ThreeDatitzPlots}. In order to produce these plots, the excitation energy of the decaying $^{12}$C was first calculated. This was first done by calculating the momentum and kinetic energy of the decaying $^{12}$C as

\begin{eqnarray}
\label{eq:CarbonP}
\vec{P}_{\textrm{$^{12}$C}} = \sum_{i=1}^{3} \vec{P}_{\alpha_\textrm{i}}\\
\label{eq:CarbonE}
E_{\textrm{$^{12}$C}} = \frac{\left | \vec{P}_{\textrm{$^{12}$C}} \right |^2}{2M_{\textrm{$^{12}$C}}}.
\end{eqnarray}

\begin{figure}[H]
\centering
\includegraphics[scale=0.9]{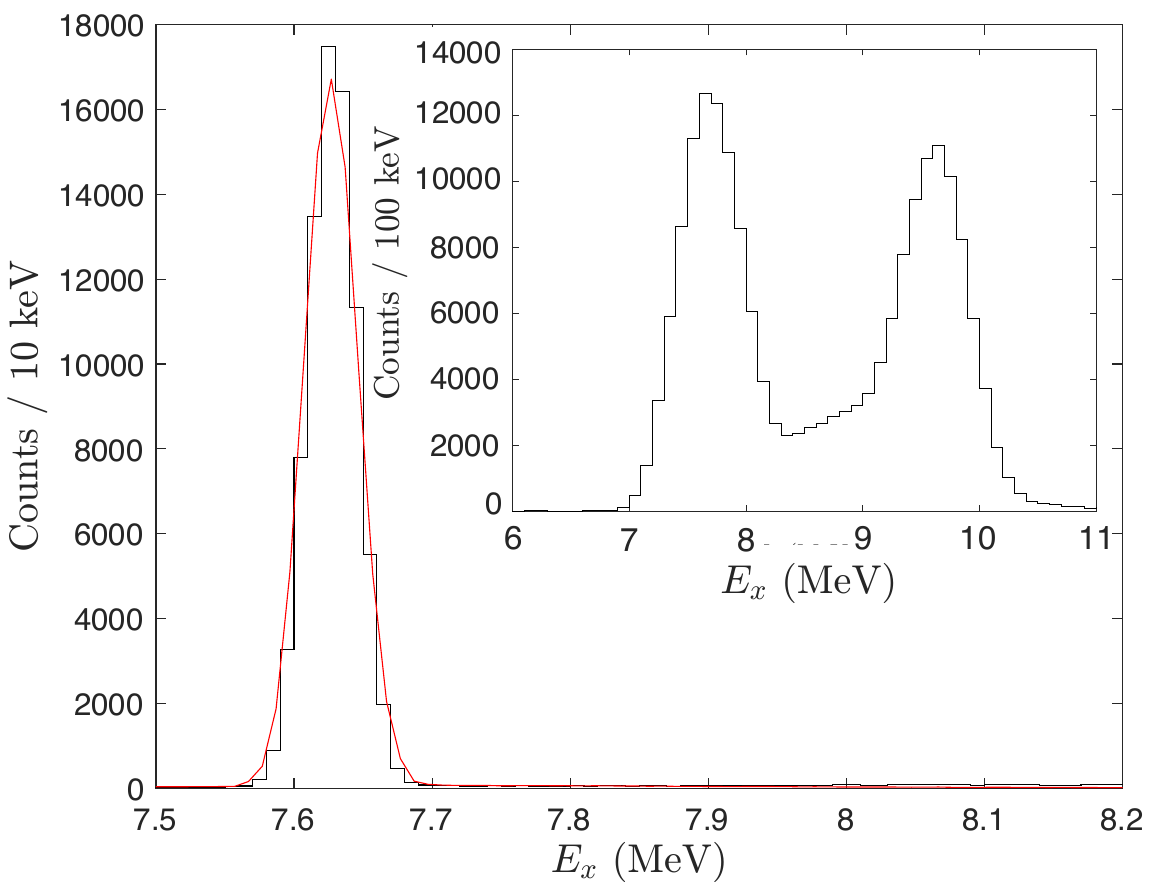}
\caption{Excitation spectra of $^{12}$C. Main figure: Complete kinematics data. Inset: Data where one $\alpha$-particle has been reconstructed. The reconstructed data show a poorer excitation energy resolution. The difference in populations of the 7.65 MeV 0$^+_2$ and 9.64 MeV 3$^-_1$ are attributed to differences in the detector acceptance for each of the two multiplicities.}
\label{fig:ExSpec}
\end{figure}

\noindent Then, the $^{12}$C excitation energy was calculated as

\begin{eqnarray}
\label{eq:CarbonEx}
E_{x} = \sum_{i=1}^{3} E_{\alpha_\textrm{i}} - E_{\textrm{$^{12}$C}} - Q.
\end{eqnarray}

\noindent where $Q$ is the $Q$-value for the decay ($-$7.27 MeV). The resulting excitation energy plots are shown in Figure \ref{fig:ExSpec}. The data corresponding to where all final state $\alpha$-particles are detected directly forms the main plot, and the data where one of the $\alpha$-particles was reconstructed is shown as the inset. As can be seen, the \emph{complete kinematics} data show a considerably better excitation energy resolution than when one particle is reconstructed. A 1.5$\sigma$ software cut was placed on the Hoyle state peak at $\approx$ 7.65 MeV. This allowed us to focus on decays from the Hoyle state, with some background from higher energy states.\\

From there, the relative energies of the three break-up $\alpha$-particles were examined in the $^{12}$C centre-of-mass frame using Dalitz plots. The construction of Dalitz plots is discussed in detail in the references \cite{Smith9BeDalitz,BishopDalitz}. In short, the symmetric Dalitz plot provides a way to draw three quantities on a two-dimensional plot, provided that the three quantities sum to a known value. In this example, the fractional energies of the three $\alpha$-particles are plotted, meaning that they must sum to unity. Since, in the sequential decay, the first emitted $\alpha$-particle carries away a fixed fraction of the decay energy ($\approx$~50\%), decays of this kind appear on a triangular locus. The direct decay, on the other hand, appears as a background to this triangle.

For reference, the Dalitz plot obtained when all four final state $\alpha$-particles are detected directly is shown in Figure \ref{fig:ThreeDatitzPlots} a). Excellent resolution is obtained and events corresponding to sequential and direct decay can be clearly distinguished. The Dalitz plot obtained when one of the $\alpha$-particles is reconstructed is shown in Figure \ref{fig:ThreeDatitzPlots} b). This has a relatively poor resolution, meaning that contributions from sequential and the rare direct decay cannot be easily differentiated. Finally, Figure \ref{fig:ThreeDatitzPlots} c) shows the same data after kinematic fitting was applied. A significant improvement in the resolution is obtained.

The \texttt{FUNKI\_FIT.F} \texttt{FORTRAN} code \cite{FUNKIFITfortran} was used for the kinematic fitting and utilised two constraint equations

\begin{eqnarray}
\label{eq:Constraint1}
\sum_{i=1}^{4} E_{\textrm{i}} - Q - E_{\textrm{beam}} = 0\\
\label{eq:Constraint2}
\sum_{i=1}^{3} E_{\alpha_\textrm{i}} - E_{\textrm{$^{12}$C}} - Q - 7.6542 \hspace{0.1cm} \textrm{MeV} = 0.
\end{eqnarray}

\noindent where $Q$ = $-$ 7.27 MeV and $E_{\textrm{$^{12}$C}}$ is calculated as shown in equations \ref{eq:CarbonP} and \ref{eq:CarbonE}. Equation \ref{eq:Constraint1} provides the constraint that the total energy of the system is conserved during the reaction. Equation \ref{eq:Constraint2} provides the constraint that the three $\alpha$-particles emerging from $^{12}$C have resulted from the decay of a state with the exact energy of the Hoyle State (7.6542 MeV). Since total momentum conservation was used to reconstruct the momentum of the missing $\alpha$-particle, this constraint is already met so cannot be utilised further in the kinematic fitting.\\

The width of the Dalitz plot distribution reduced by a factor of $\approx$ 1.5 through kinematic fitting. This is demonstrated in the left panel of figure \ref{fig:FoldedDatitzPlots}, which shows a particular projection of the Dalitz plot. To obtain this projection, the Dalitz plots of figure \ref{fig:ThreeDatitzPlots} were folded along each of the lines of symmetry, defined by the three axes, $\epsilon_1$, $\epsilon_2$ and $\epsilon_3$. After this procedure, the data occupy a single sextant of the Dalitz plot. The data may then be projected onto the remaining $\epsilon$ axis, in order to examine the width of the triangular locus. The left panel of figure \ref{fig:FoldedDatitzPlots} shows this projection before and after kinematic fitting, illustrating the improvement in resolution.

\begin{figure}[H]
\centering
\includegraphics[scale=0.39]{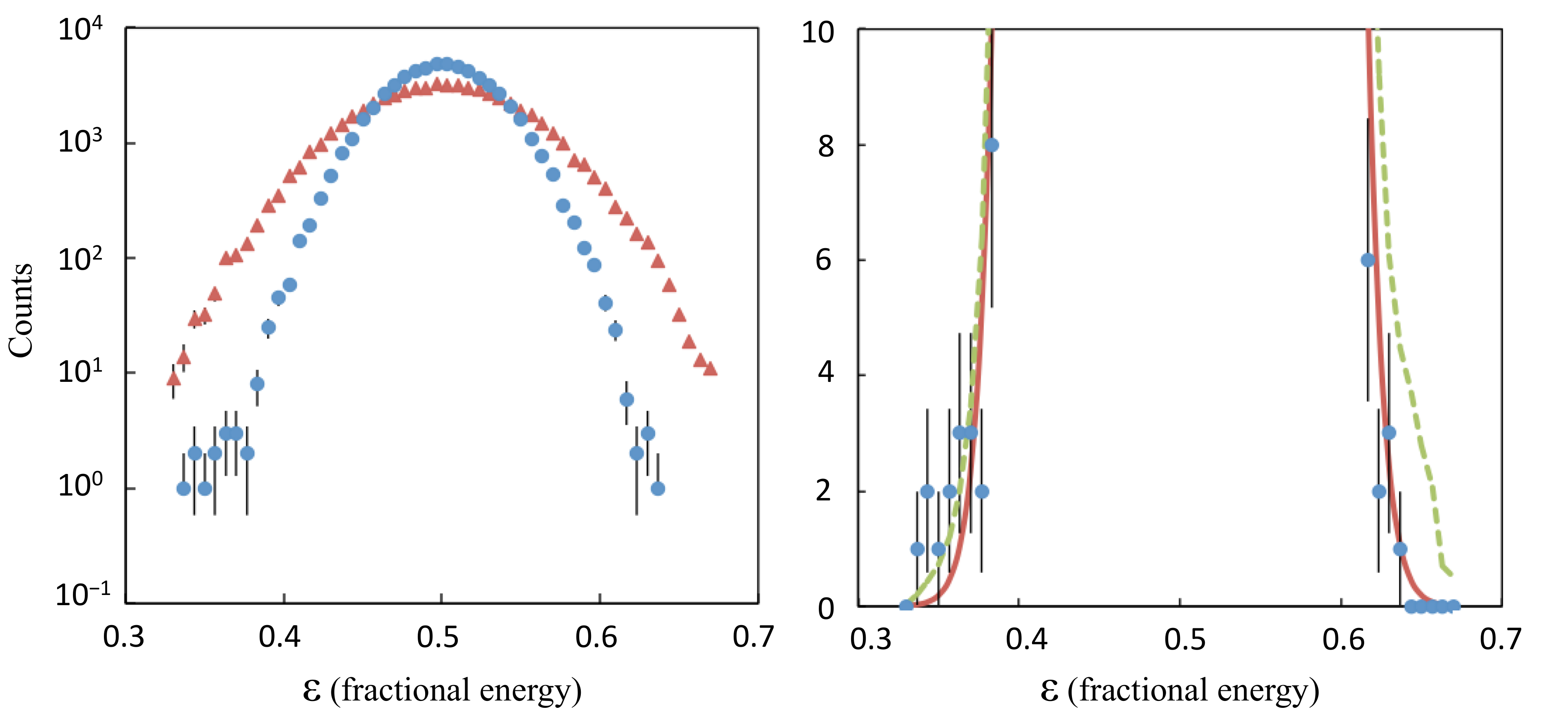}
\caption{One-dimensional projections of the Dalitz plots showing the relative fractional energies of $\alpha$-particles emitted from $^{12}$C. Left panel: Projections of the Dalitz plots before (red triangles) and after (blue circles) kinematic fitting. Right panel: Fits of simulated data to the projected Dalitz plot after kinematic fitting. The red, solid line shows a 100\% sequential decay, and the green, dashed line includes a 0.3\% direct decay contribution.}
\label{fig:FoldedDatitzPlots}
\end{figure}

This improvement means that the relative contributions of sequential and direct decay could be more clearly constrained. To quantify this, Monte-Carlo simulations of sequential and direct decay events were generated. Details of the Monte-Carlo code can be found in references \cite{RES81} and \cite{RES82}. To demonstrate how each decay type appears on the Dalitz plot, the simulated sequential and direct decays (for complete kinematics, high resolution events) are shown as the folded Dalitz plots in figure \ref{fig:FoldedDatitzPlots}. To evaluate the relative sequential and direct decay contributions, the Monte-Carlo data were fitted to the experimental data. The right panel of figure \ref{fig:FoldedDatitzPlots} shows the tails of the Dalitz plot (the area most sensitive to direct decay) for the kinematic fitted data. The solid line shows the simulation of a 100\% sequential decay, which gives a $\chi^2$ per degree of freedom of 1.08, close to the 50\% confidence level (C.L.), indicating a good fit. Once the direct decay contribution was increased to 0.2\%, the $\chi^2$ value increased beyond the 99\% C.L. Thus an experimental upper limit of 0.2\% was assigned to the rare direct decay mode. In contrast, the same analysis, applied to the pre-kinematic fitted data with poorer resolution, obtained an upper limit of 1.4\% for the direct decay mode. Thus, the kinematic fitting improves the sensitivity to the direct decay by almost an order of magnitude.

\begin{figure}[H]
\centering
\includegraphics[scale=0.78]{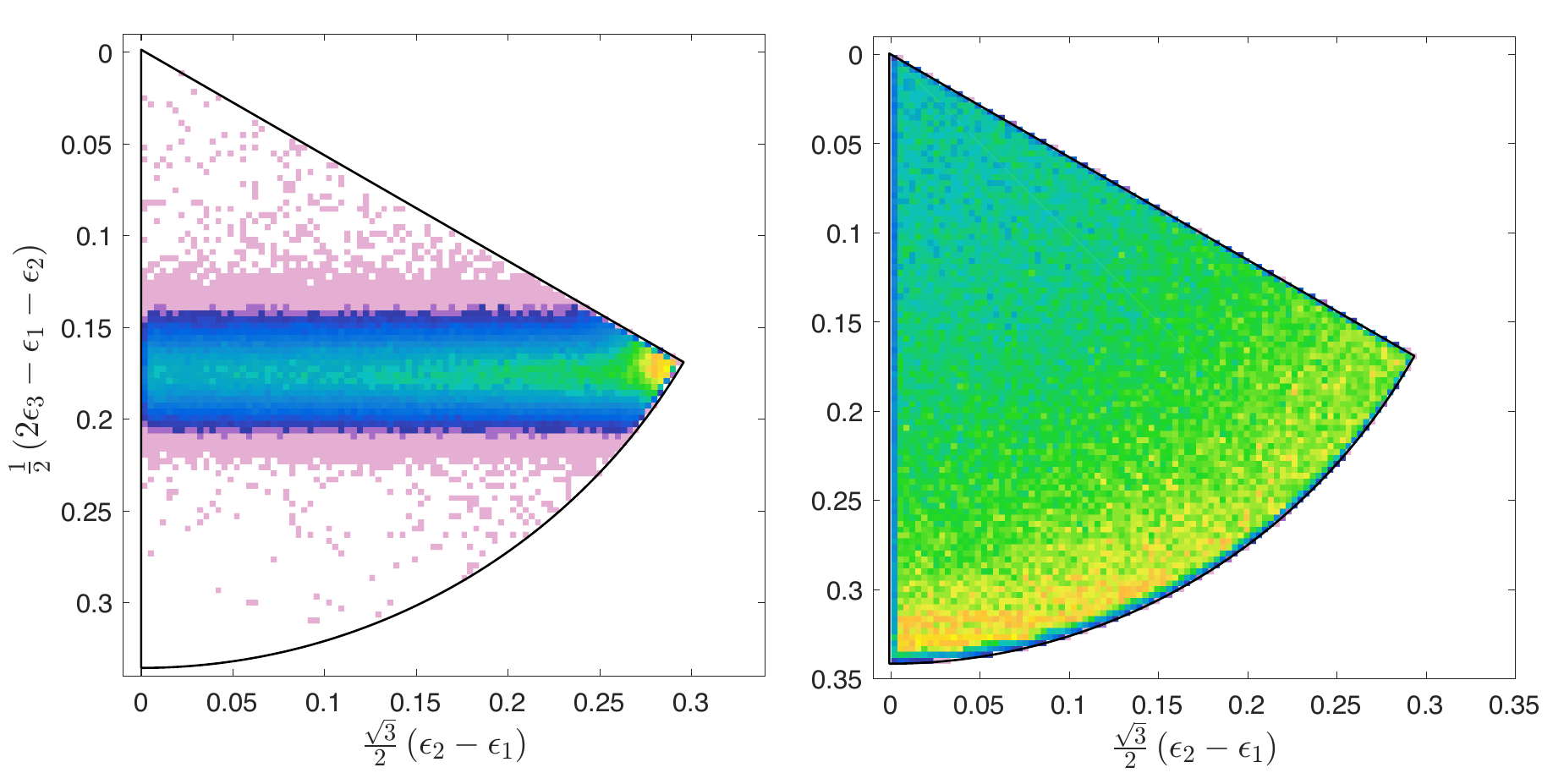}
\caption{Folded Dalitz plots of simulated, complete kinematics data. Left panel: sequential decay. Right panel: direct decay. Note that the direct decay mechanism was modelled as an equal probabilities decay to anywhere in the available phase space, corresponding to a uniform filling of the Dalitz plot in the right panel. The slight non-uniformity of the simulation is due to the detector acceptance.}
\label{fig:FoldedDatitzPlots}
\end{figure}

Dalitz plot analyses are also commonly used when studying the decays of other borromean light nuclei, such as the neutron-rich $^{14}$Be \cite{Marquesdalitzplot}, $^{18}$C and $^{20}$O \cite{Marquesdalitzplot2}. These are predicted to have neutron halo structures that have previously been investigated through examining the relative energies of the emitted neutrons. Despite recent neutron detector upgrades, the neutron momenta are notoriously difficult to measure and could benefit from the presented kinematic fitting routine.

\subsection{Application to TPCs}
\label{sec:intro}
TPCs have seen a recent boom in their use in nuclear physics experiments \cite{TPC1,TPC2}. One of their primary areas of application is in the use of the Thick Target Inverse Kinematics (TTIK) technique \cite{TTIK}. This involves a beam passing through a gas target medium and losing energy as it does so. This causes the beam + gas system to sweep out a range of centre-of-mass energies. Therefore, any compound nucleus reaction inside the gas can have its centre-of-mass energy identified from the interaction location. In this way, an excitation function can be measured by counting the number of interactions as the beam passes through the gas.\\

This works well but there are several limiting factors. Firstly, for beams that are unstable, and therefore must be created using a recoil separator or the ISOL technique, by the time the beam has entered the TPC, the beam energy spread can be very large ($\approx 5\%$ in the case of a recoil separator). This has a dramatic effect on the centre-of-mass energy resolution as the vertex location no longer well-defines the centre-of-mass energy. As an alternative, an accurate energy and angle measurement of the recoil products of a reaction can be used instead to calculate the centre-of-mass energy. Furthermore, for a range of situations, the recoil products stop inside the gas and cannot be well measured by an exterior silicon detector; this also limits the energy resolution. Additionally, the stopping power for the recoil + gas system may not be well known or prone to systematic differences, which can also degrade the resolution. A final additional experimental hurdle corresponds to the position resolution and, therefore, the momentum vector, for light recoil products. For high energy protons, the energy loss in the TPC region may fall below the threshold for detection and therefore can only be detected in silicon detectors, which may not be well pixellated.\\

To rectify this, one can constrain the system by measuring the incoming beam track and the two recoil product tracks (which have correspondingly large uncertainties) and apply kinematic fitting constraints to improve the reconstructed energy resolution. Here, the parameters are that of the momentum for the beam (${p_b}$), light (${p_l}$) and heavy (${p_h}$) recoil products and their corresponding polar and azimuthal angles ($\theta,\phi$) just before/after the interaction with Q-value, $Q$. The following constraints can then be applied:
\begin{eqnarray}
\label{eq:TPCconstraint1}
\vec{p_{b}} = \vec{p_{h}} + \vec{p_{l}} \\
\label{eq:TPCconstraint2}
\frac{{p_b}^2}{2 m_b} = \frac{{p_h}^2}{2 m_h} + \frac{{p_l}^2}{2 m_l} - Q.
\end{eqnarray}

In this way, one can parameterise the energy spread of the incoming beam by setting the initial $p_b$ based on the expected energy loss of the beam to the measured interaction vertex with a variance given by the beam energy resolution, or by using the position resolution with which one can identify the reaction vertex. After these parameters have been constrained, the centre-of-mass energy $E_{CM}$ is then simply:

\begin{eqnarray}
E_{CM} = \frac{{p_b}^2}{2 m_b}\left ( \frac{m_{t}}{m_{t}+m_{b}} \right )
\end{eqnarray}

\noindent with $m_t$ being the mass of the target and $m_b$ being the mass of the beam.\\

It is worth noting that a similar result can be obtained by using a combination of the recoil product energies and the incoming beam energies without the use of kinematic fitting. However, the kinematic fitting not only deals with these multiple sources of errors well on an event-by-event basis but also provides a similar improvement to the angular resolution, which is not possible without the use of this technique. This angular resolution is particularly important in TTIK and transfer experiments where the spin information is encoded in the angular dependencies of the measured reactions.

\subsection{Simulated TPC data}
In order to benchmark this approach, a Geant4 simulation was used that simulates the detector response of the TexAT (Texas Active Target) TPC \cite{TexAT2}. The detailed simulation includes the full detector geometry, segmentation of the readout plane, and diffusion of electrons as they drift through the TPC. The interaction simulated was that of $^{10}\mathrm{C}(\alpha,\alpha)$ elastic scattering. An isolated resonance was simulated at $E_{CM} = 2.5$ MeV and the incident $^{10}\mathrm{C}$ was given an energy of exactly 18 MeV upon entering the chamber, filled with 600 Torr of helium gas. Since the initial energy of the beam is known, the loss of its energy as it passes through the detector will mean that the population of the 2.5 MeV resonance will occur at a fixed position in the detector.\\

The excitation energy resolution is therefore partly defined by the accuracy by which the position of the interaction point can be determined. The 3D tracks were then reconstructed from the waveforms generated from the simulation, using the same analysis code that is used to analyse real experimental data. The segmentation of TexAT is 1.75 x 3.5 mm in the central (beam) region and has 1.75 mm pitch strips and chains either side of the central region. Therefore, this is indicative of the track resolution one can achieve.\\

\begin{figure}[H]
\centering
\includegraphics[scale=0.65]{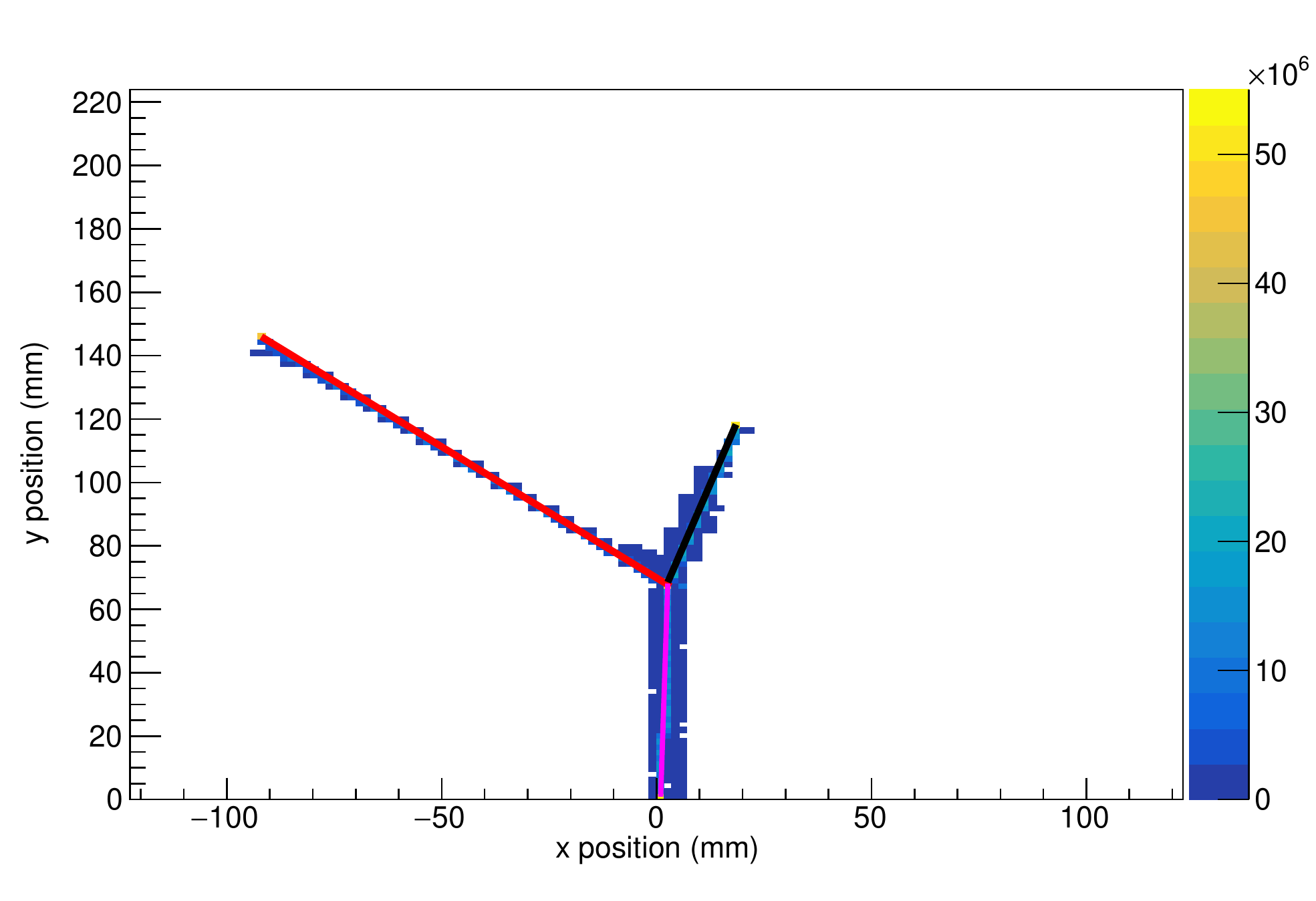}
\caption{XY projection of a simulated $^{10}\mathrm{C}(\alpha,\alpha)$ reaction track in TexAT, reconstructed from a Monte Carlo generated event where the histogram colour represents the energy deposited in the Micromegas of TexAT. The magenta line is the fitted incoming track for the beam, the red line is the fitted light recoil ($\alpha$) track and the black line is the fitted heavy recoil track ($^{10}\mathrm{C}$).}
\label{fig:track}
\end{figure}

\begin{figure}[H]
\centering
\includegraphics[scale=0.65]{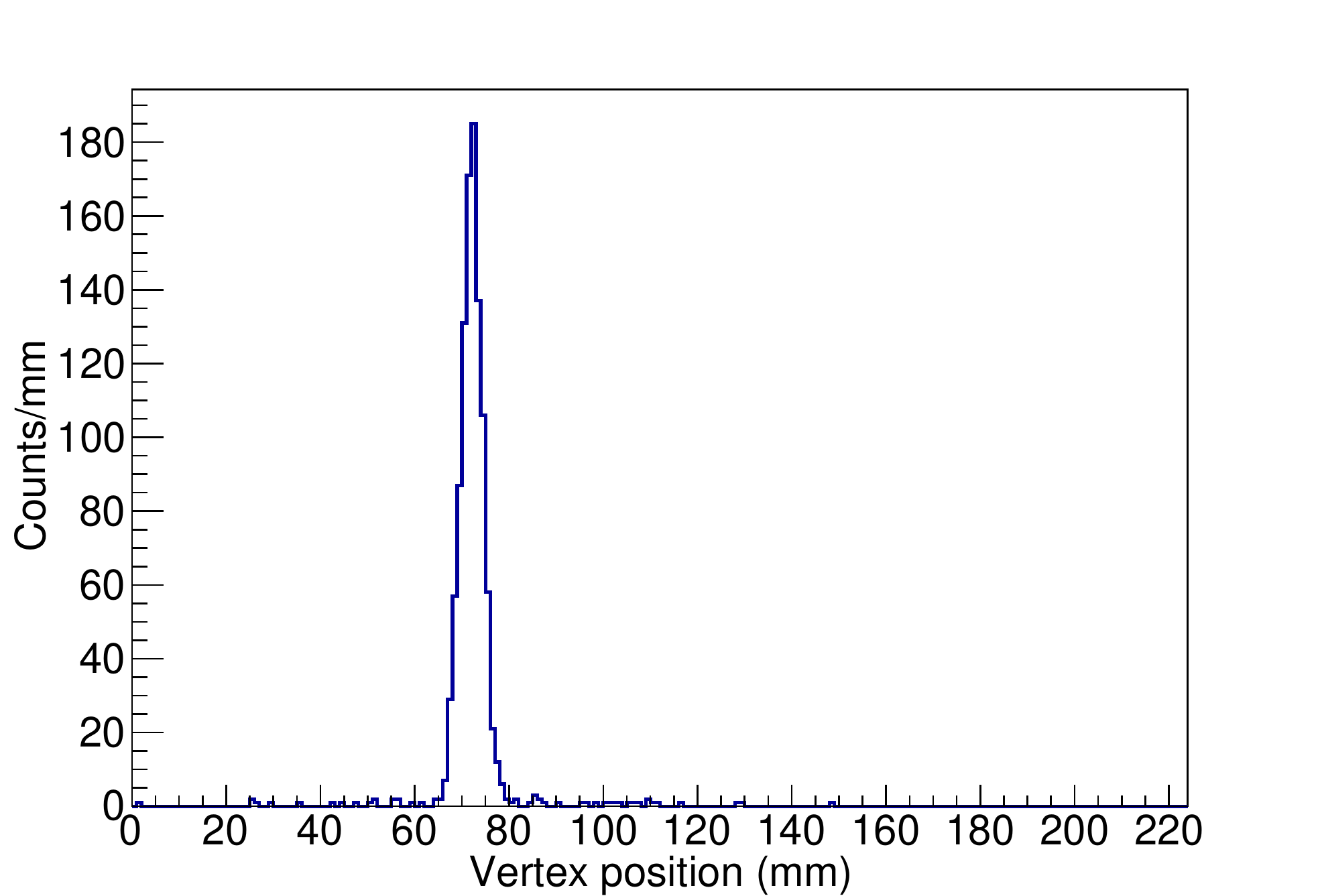}
\caption{Reconstructed vertex positions along the beam direction of $^{10}\mathrm{C}(\alpha,\alpha)$ reaction tracks in TexAT, reconstructed from Monte Carlo generated events with no beam energy spread. The main peak can be clearly be seen and has a width of $\sigma$ = 2.5 mm. The events away from this peak correspond to misidentified vertex locations due to the limitations of the TexAT segmentation, vertex fitting and Hough transform algorithm. They mainly correspond to small angle scattering events where the light and heavy recoils are hard to distinguish.}
\label{fig:vertex}
\end{figure}

To find the interaction vertex, the incoming beam and the two recoil tracks were found using a Hough transform and then fitted using the least-squares distance approach to give the interaction vertex, the trajectory of the incoming beam and the end points of the two recoil particles. This least-squared approach also included terms to avoid over/underfitting the length of the arms so that they terminated at the end of the reconstructed tracks. An example of this fit can be seen in Figure~\ref{fig:track}. This fit also gives the associated covariance matrices that can be used for the later kinematic fitting using the ROOT version of \texttt{FUNKI\_FIT} \cite{FUNKIFITroot}. A histogram of vertex positions, calculated for all simulated events, is shown in Figure~\ref{fig:vertex}. An average vertex reconstruction error of 2.5 mm was achieved along the beam direction corresponding to roughly 250~keV energy resolution (70~keV centre-of-mass energy resolution).

\begin{figure}[H]
\centering
\includegraphics[scale=0.65]{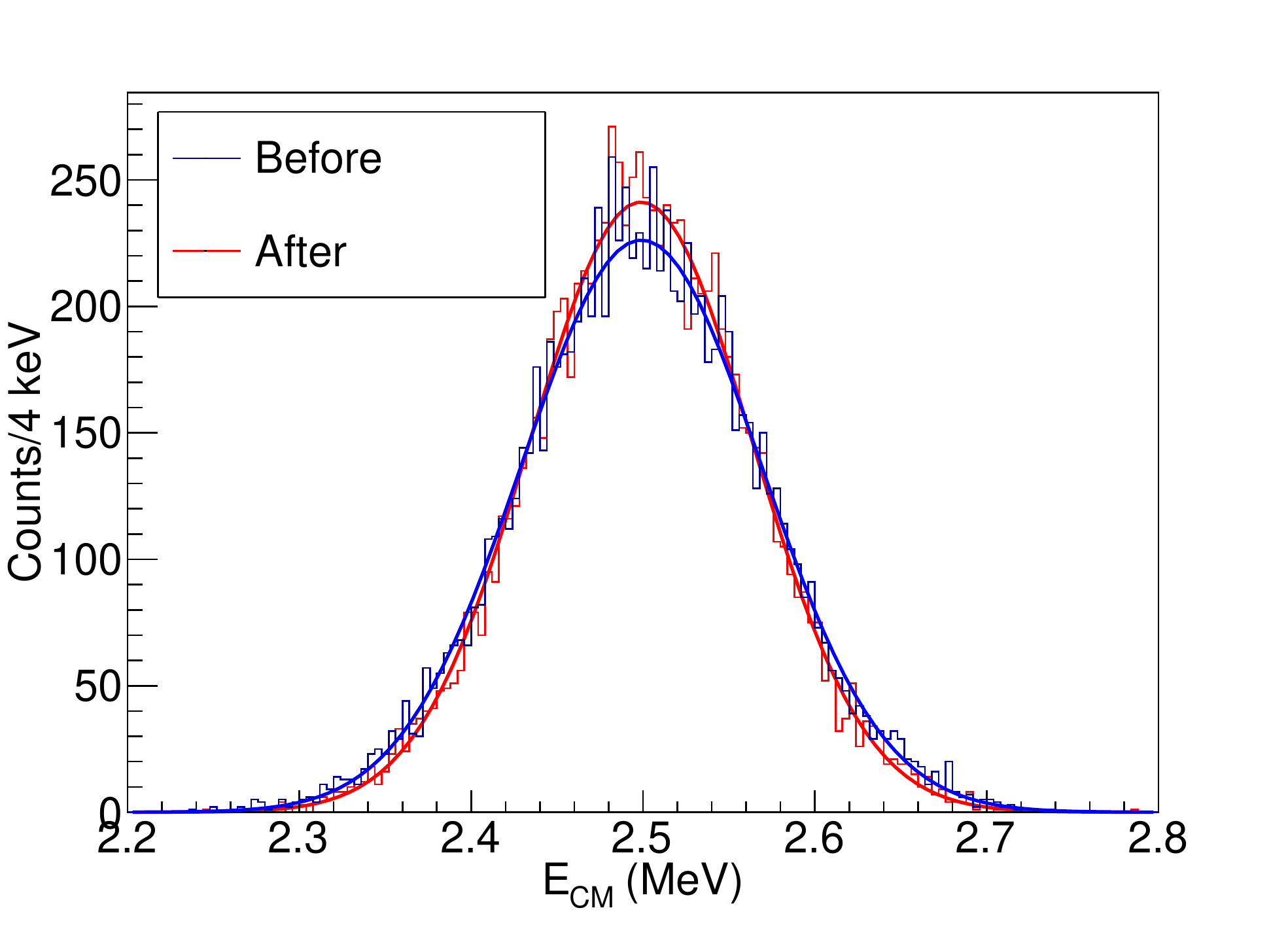}
\caption{Centre-of-mass energy as calculated from the energies of the recoil products for a $E_{CM} ~=~2.5$~MeV resonance in the $^{10}\mathrm{C}(\alpha,\alpha)$ channel.}
\label{fig:ECMKF}
\end{figure}

To obtain an independent measure of the centre-of-mass energy, one must use the energies of the recoil products, which were determined in the simulated data by measuring the range of the reaction products in the gas, as determined from the fit. A measurement of the energies from the ranges has a dramatically worse resolution that via a direct energy measurement with a silicon detector. The energies and polar angle of each particle (the beam, light $\alpha$-particle and heavy $^{12}$C recoil) are then used to calculate the vector momenta used in constraint equations \ref{eq:TPCconstraint1} and \ref{eq:TPCconstraint2}. Despite this large error in centre-of-mass energy from the recoil products ($\sigma = $ 175 keV), the kinematic fitting technique still provides an improvement, which can be seen in Figure~\ref{fig:ECMKF}. This was obtained using the formulation above in conjunction with the centre-of-mass measurement from the beam which has a superior energy resolution in this case.\\

\begin{figure}[H]
\centering
\includegraphics[scale=0.65]{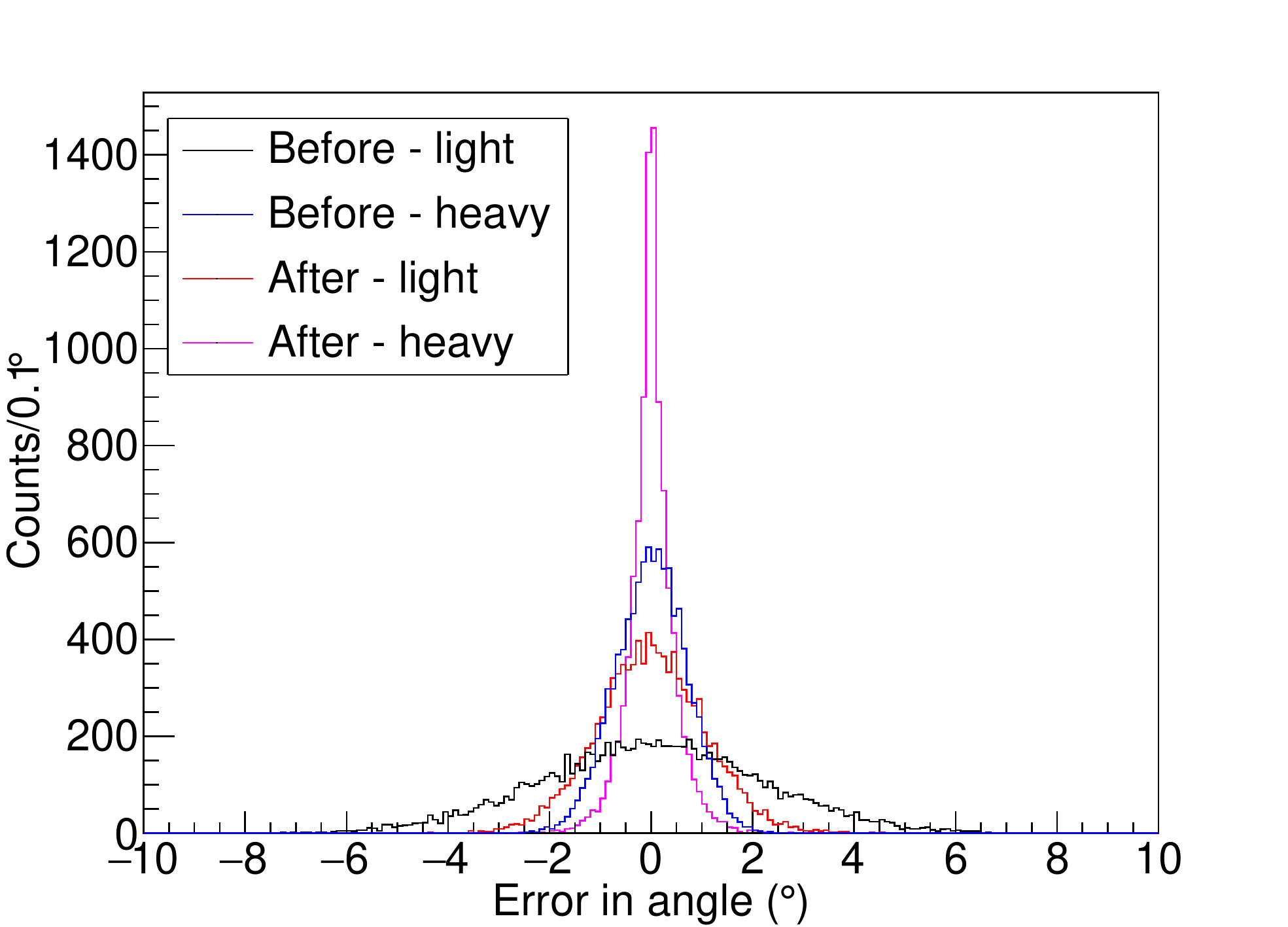}
\caption{Uncertainty in polar angle for the light and heavy particles calculated before and after the kinematic fitting process in the $^{10}\mathrm{C}(\alpha,\alpha)$ channel. The effect of the kinematic fitting can be seen as a dramatic improvement in the angular resolution for both the light and heavy products.}
\label{fig:TexATangle}
\end{figure}

\begin{figure}[H]
\centering
\includegraphics[scale=0.65]{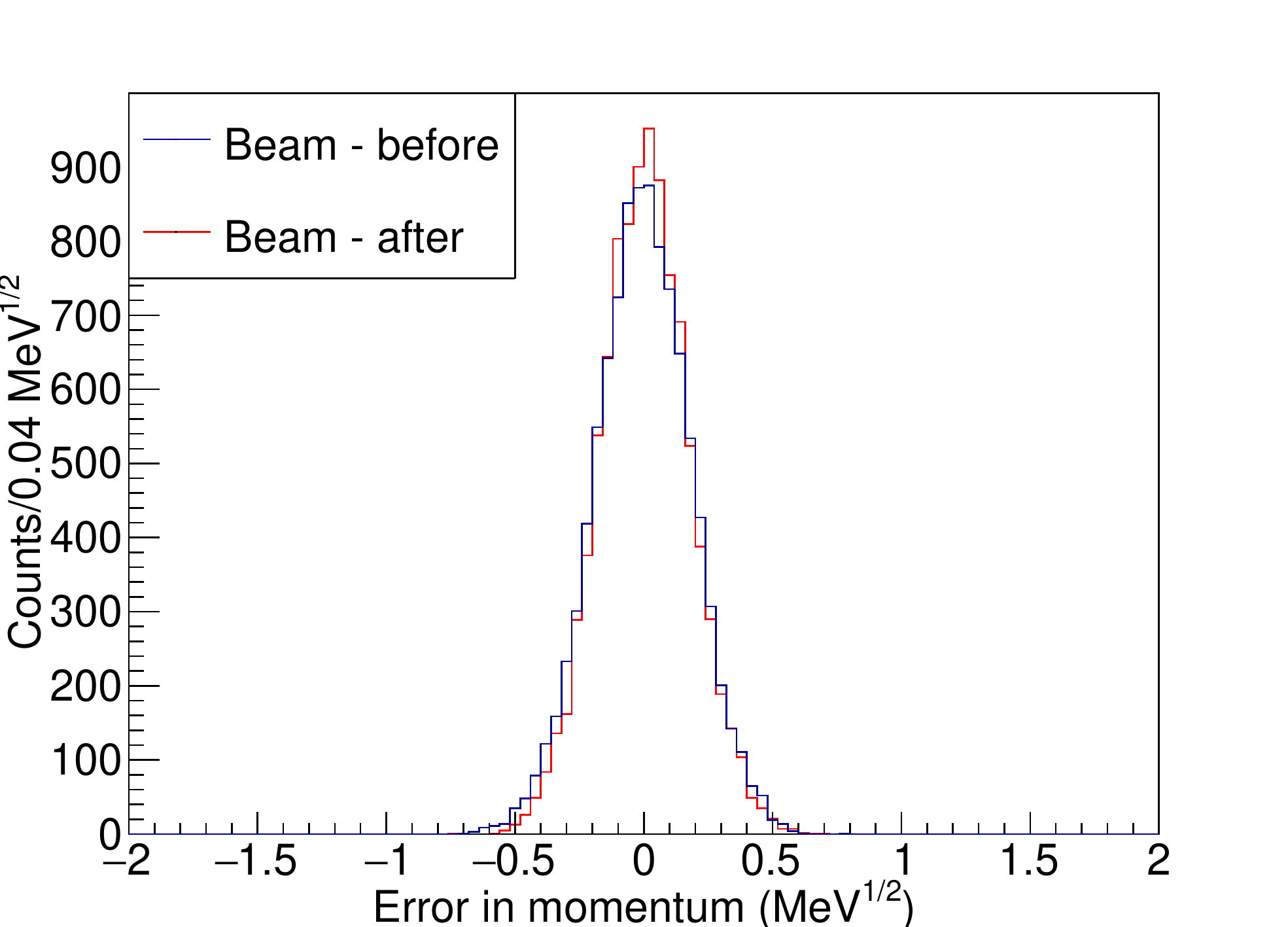}
\caption{Uncertainty in the calculated beam momentum before and after kinematic fitting.}
\label{fig:BeamP}
\end{figure}

\begin{figure}[H]
\centering
\includegraphics[scale=0.65]{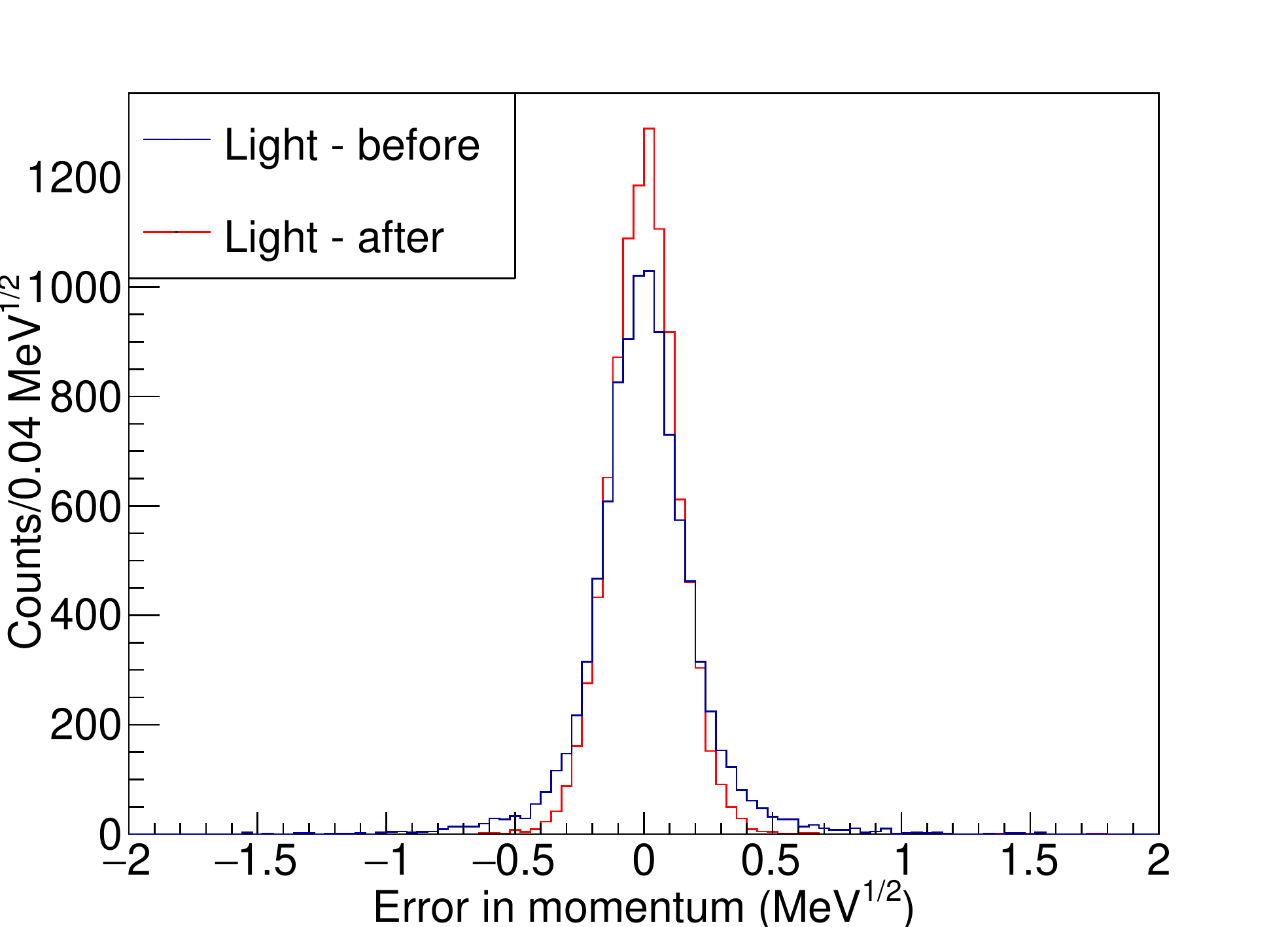}
\caption{Uncertainty in the calculated momentum of the light $\alpha$-particle, in the $^{10}\mathrm{C}(\alpha,\alpha)$ channel, before and after kinematic fitting.}
\label{fig:LightP}
\end{figure}

\begin{figure}[H]
\centering
\includegraphics[scale=0.65]{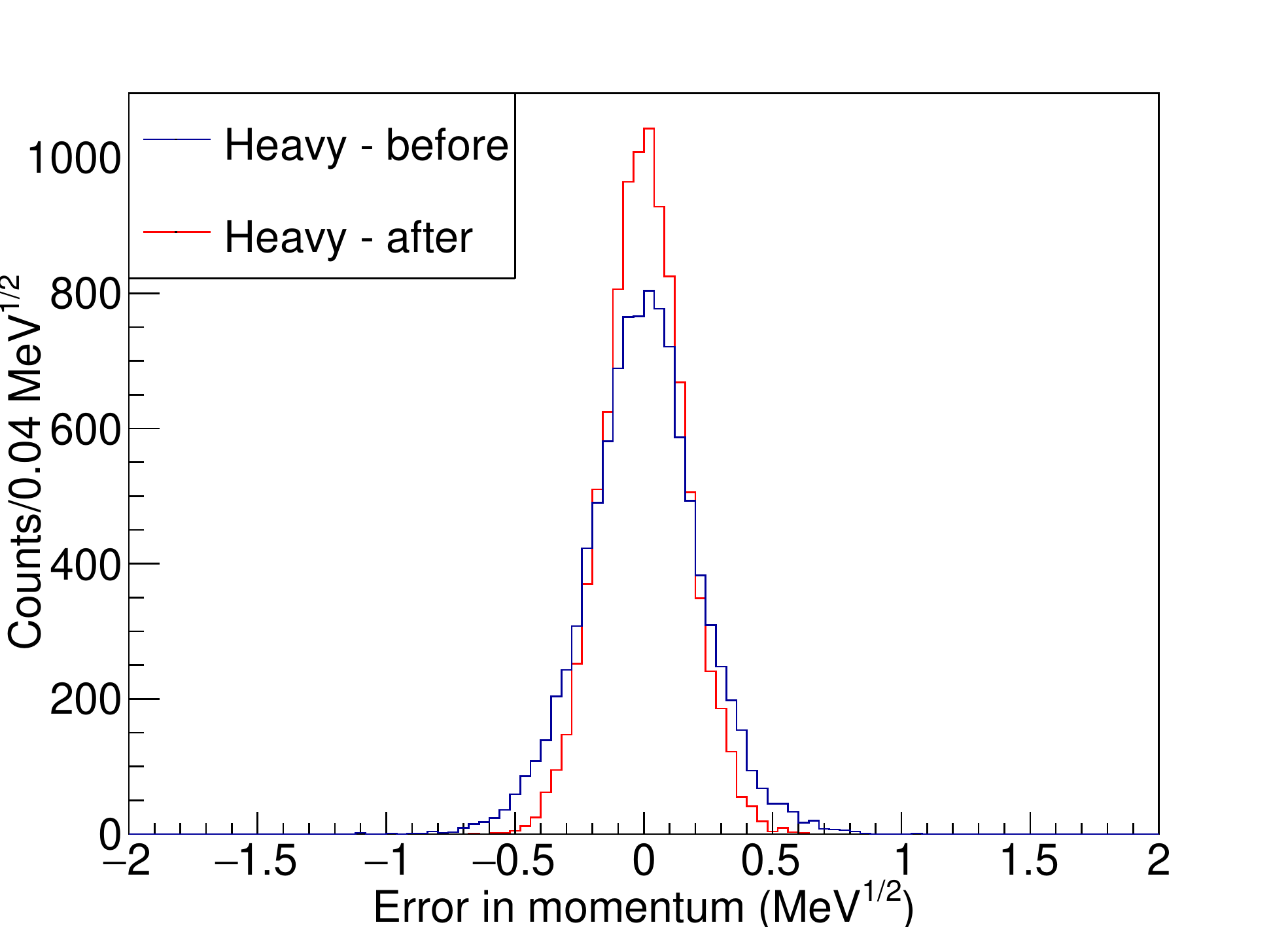}
\caption{Uncertainty in the calculated momentum of the heavy $^{10}$C, in the $^{10}\mathrm{C}(\alpha,\alpha)$ channel, before and after kinematic fitting.}
\label{fig:HeavyP}
\end{figure}

The resolution can be seen to change from $\sigma$ = 70 keV to $\sigma_{KF}$ = 65 keV. This is consistent with taking the weighted mean ($w_i = \sigma_i ^{-2}$) of the centre-of-mass energy resolution from the beam and the recoil products ($\sigma$ = 66 keV). Most dramatically, the kinematic fitting technique also improves the angular resolution where the deviation from the true value can be seen in Figure~\ref{fig:TexATangle}. Kinematic fitting improves the angular error, $\sigma \theta$, (in the lab frame) from 2.2$^{\circ}$ and 0.7$^{\circ}$ for the light and heavy recoil respectively to 1.0$^{\circ}$ and 0.4$^{\circ}$. As mentioned, this improvement is essential for the extraction of spin-parity information in TTIK measurements from R-Matrix theory. The light recoil has a larger error as it has an overall larger scattering angle. For completeness, although they carry less physical significance, the uncertainties in the momenta of the beam, light $\alpha$-particle, and heavy $^{10}$C are shown explicitly in figures \ref{fig:BeamP}$-$\ref{fig:HeavyP}, since these parameters are used in the constraint equations.

\par

This approach is useful not just in the shown case, where the errors on the particle momenta are large, but also in situations where the recoil particles are measured in a silicon detector. For detector configurations that require a large solid-angle coverage of silicon shells, the angular granularity of these detectors may be very large and therefore there is a large momentum uncertainty. Furthermore, this case emphasises the situation whereby the beam energy resolution is smaller than that from measuring the reaction products. However, this technique is equally as applicable in the case of a poor beam energy resolution, such as when using radioactive beams as discussed above.

\section{Discussion}

The present study demonstrates, for the first time, the usefulness of kinematic fitting in low energy nuclear physics across a range of detector technologies, experimental methods, and scientific aims.\\

In charged-particle spectroscopy using silicon detectors, a significant improvement was seen in the fractional energy resolution of the $\alpha$-particles emitted during the decay of $^{12}$C. This allowed the various decay types of $^{12}$C to be differentiated more clearly. This success has consequences for studying the break-up of other borromean light nuclei such as $^9$Be. When studying $^9$Be, typically only the $\alpha$-particles resulting from the $^9$Be $\rightarrow \alpha + \alpha + n$ decay are measured. The neutron is undetected and reconstructed as missing mass. In reference \cite{Smith9BeDalitz}, the $\alpha$-$\alpha$-$n$ relative energies were examined using Dalitz plots, with a resolution that could be improved using kinematic fitting. Likewise, the study of more neutron rich borromean nuclei, such as $^{14}$Be \cite{Marquesdalitzplot}, $^{18}$C and $^{20}$O \cite{Marquesdalitzplot2} may reconstruct missing neutrons or detect them directly. Such neutron detectors typically have poor energy and position resolution, so these analyses could be improved using kinematic fitting. With the growing popularity of radioactive beam facilities, where the behaviour of more neutron rich nuclei are opened up to exploration, kinematic fitting could play an important role.\\

As previously noted, the use of TPCs is increasing in nuclear physics experiments \cite{TPC1,TPC2}. In this paper, we have demonstrated, for the first time, that one of the primary applications of TPCs, Thick Target Inverse Kinematics (TTIK) measurements, are dramatically improved using kinematic fitting. We record a modest improvement in the excitation energy resolution and large improvements in angular resolution. For fitting spectra obtained with TTIK, using the $R$ matrix method, detailed angular measurements are required in order to constrain the angular momenta of states in the spectra. Therefore, such improvements in angular resolution are of great significance.

\authorcontributions{Conceptualization, R.~Smith and J.~Bishop; methodology, R.~Smith and J.~Bishop; software, R.~Smith and J.~Bishop; validation, R.~Smith and J.~Bishop; formal analysis, R.~Smith and J.~Bishop; investigation, R.~Smith and J.~Bishop; resources, R.~Smith and J.~Bishop; data curation, R.~Smith and J.~Bishop; writing--original draft preparation, R.~Smith and J.~Bishop; writing--review and editing, R.~Smith and J.~Bishop; visualization, R.~Smith and J.~Bishop; supervision, R.~Smith and J.~Bishop; project administration, R.~Smith and J.~Bishop; funding acquisition, R.~Smith and J.~Bishop.}

\funding{This research was funded by the United Kingdom Science and
Technology Facilities Council (STFC) under grant number ST/L005751/1 and the US Dept. of Energy Federal Award Number DE-FG02-94ER40870.}

\acknowledgments{The assistance of the staff at the University of Birmingham MC40 Cyclotron is
gratefully acknowledged. Further thanks are given to C. Wheldon for his input to discussions about the initial use of kinematic fitting. JB would also like to thank the members of the TexAT collaboration especially those involved in developing the GEANT4 framework.}

\conflictsofinterest{The authors declare no conflict of interest.}




\reftitle{References}





\end{document}